\documentclass[10pt]{article}
\date{}
\title{Qualms regarding \lq\lq Non-extensive Hamilton systems follow Boltzmann's
principle not Tsallis statistics--phase transitions, second law of Thermodynamics\rq\rq\
by D. H. E. Gross, \emph{Physica A} \textbf{305} (2002) 99}
   \author{B. H. Lavenda$^1$ and J. Dunning-Davies$^2$\\
$^1$Universit\`a degli Studi  Camerino 62032 (MC) Italy;\\ email: bernard.lavenda@unicam.it\\
$^2$ Department of Physics, University of Hull, Hull HU6
7RX\\ England; email: j.dunning-davies@hull.ac.uk}
\usepackage{eufrak}
\newcommand{\half}{\mbox{\small$\frac{1}{2}$}}

\begin{document}
\maketitle
\begin{abstract} 
A number of elementary misconceptions on the part of the author are corrected.
\end{abstract}
\flushbottom
The author contends that all that is needed in statistical mechanics is Boltzmann's 
principle \cite{Khinchin}
\begin{equation}
S=k\ln W\label{eq:Boltzmann}
\end{equation}
where $S$ is the entropy in a \emph{microcanonical\/} ensemble in the space $x_1,\ldots,
x_m$, and $W\;dx_1,\cdots dx_m$ is the total number of microstates in the range
$dx_1\cdots dx_m$. The author boldy states \lq\lq I take the principle as \emph{the
fundamental, generic definition of entropy.\/}\rq\rq\ In particular, the author discards 
the fundamental property of concavity \cite{LDD}.\par
Yet, his stability criterion is determined by the positive definiteness of the Hessian
\begin{equation}
\left(\frac{\partial^2S}{\partial E^2}\right)_{N,V}
\left(\frac{\partial^2S}{\partial N^2}\right)_{E,V}-
\left(\frac{\partial^2S}{\partial E\partial N}\right)_V\;\;\;
\left(=\lambda_1\lambda_2\right)\ge0. \label{eq:Hess}
\end{equation}
Had he considered the full set of independent, extensive variables, $S=S(E,V,N)$, 
 he would have found that the Hessian vanishes \cite{Lavenda91}. However, (\ref{eq:Hess}) is
not sufficient to guarantee that a twice differentiable  function should be concave.
It is also necessary that
\begin{equation}
\left(\frac{\partial^2S}{\partial E^2}\right)_{N,V}
\le0
\;\;\;\;\;\;\;\;\;\;\;\;\;\;\;
\mbox{and}
\;\;\;\;\;\;\;\;\;\;\;\;\;\;\;
\left(\frac{\partial^2S}{\partial N^2}\right)_{E,V}
\le0. \label{eq:Hardy}
\end{equation}
Taken together, (\ref{eq:Hess}) and (\ref{eq:Hardy})  imply that
the quadratic form
\begin{equation}
\left(\frac{\partial^2S}{\partial E^2}\right)_{N,V}
u^2+
2\left(\frac{\partial^2S}{\partial E\partial N}\right)_Vuv+
\left(\frac{\partial^2S}{\partial N^2}\right)_{E,V}
v^2 \label{eq:quadratic}
\end{equation}
be negative for all $u,v$. This is a necessary and sufficient condition that $S(E,V,N)$ 
be concave in any two of its variables \cite{Hardy}. These conditions are well-known in thermodynamics 
\cite{Callen}, and show that the concavity criterion is crucial to thermodynamic 
stability.\par
The author claims that a single phase exists if 
$\lambda_1<0$ since $\lambda_1\ge\lambda_2$, and this makes the Hessian, (\ref{eq:Hess})
 positive definite. This is false since all three conditions, (\ref{eq:Hess}) and
 (\ref{eq:Hardy}), are required. According to the author, the criterion for 
a transition of first-order  with phase separation is $\lambda_1>0$. The flattening of
the entropy potential at the critical point corresponds to a failure of the  concavity 
requirements. According to the generally accepted classification of phase 
transitions \cite{Tisza}, first-order phase transitions are those in which the energy and 
volume  change discontinuously, whereas phase
transitions of the second kind manifest singularities in the derivatives of the energy and volume with respect
to the temperature and pressure,  (\emph{i.e.\/}, heat capacity at constant
volume, coefficient of thermal expansion, and isothermal compressibility),
 while the energy and volume vary continuously.\footnote{The distinction between
transitions of the second kind and those of second-order is that the latter term
implies that the compliance coefficients have discontinuities at the critical
temperature, but have finite values in both phases. In transitions of the second
kind, the compliance coefficients become singular \cite{Tisza}.}  \par
The eigenvalues of the
quadratic form, 
\[\lambda_{1,2}=\half(S_{NN}+S_{EE})\pm\half\sqrt{(S_{NN}-S_{EE})^2+4S_{NE}^2},\]
where the subscripts indicate partial derivatives, are devoid of any physical 
meaning. Rather, one must complete the square in order to obtain coefficients
of a quadratic form that are given in terms of the principal minors of the
compliance matrix \cite{Tisza}. Hence, the condition that $\lambda_1=0$ is not a 
condition for a critical point, contrary to what the author claims.\par
Gross also mentions that in the canonical ensemble the mean value of the energy is fixed 
by the temperature, $1/\beta$. Rather, it is $\beta$ which is the estimable 
parameter \cite{Mandelbrot,Lavenda91}, 
and since the canonical distribution belongs to the exponential family, any energy 
of a sample system belonging to a population that is  believed to be at uniform $\beta$,
 or the average of the energies of the sample systems
of the entire population, is a sufficient statistic to estimate $\beta$. That is to say,
 a finer division of the system does not give a better estimate of $\beta$.\par Furthermore, the 
statement that in order to agree with the microcanonical ensemble, 
$e^{-\beta E}W(E)$ must be sharp in $E$ has no meaning since $\beta$ is not defined
in a microcanonical ensemble. Recall that there are two very distinct operations
 after $W$ has been equated to the thermodynamic probability \cite{Fowler}. Firstly,  
determining the maximum of $W$ fixes the most probable state of the isolated
system \textit{by itself\/}. That is, the greatest number of complexions that correspond
to a single macroscopic state. Secondly, the system is placed in thermal contact with the 
\textit{outside world\/}, and the second law $\partial S/\partial E=1/T$ 
relates the thermodynamic entropy,  determined by the maximum
of $W$, to the temperature $T$ of the outside world.\par
Gross also asserts that the  extensivity (additivity) of the entropy is not required 
since Boltzmann's principle 
\lq\lq \emph{is all that Statistical Mechanics demands\/}\rq\rq. For a statistically
independent assemblies, the joint probability reduces to the product of the individual
probabilities
\begin{equation}
W_{12}=W_{1}W_{2}.\label{eq:W}
\end{equation}
Then by Boltzmann's principle (\ref{eq:Boltzmann}) it follows that
\begin{equation}
S_{12}=S_1+S_2. \label{eq:S}
\end{equation}
Hence, extensivity (additivity) is already incorporated in Boltzmann's principle. The 
argument is based on the fact that the thermodynamic probability $W$ is to be treated
as a true probability, which it is not. Any attempt to normalize it leads to a term which 
for all intent and purposes is unity \cite{Fowler}. Moreover, we know that the entropies of
two bodies add when surface interactions are neglected and both are at the same 
temperature. Again Fowler questions the fact that the two bodies cannot be statistically
independent because one body \lq knows\rq\ the temperature of the other. Yet, if we consider
initially one body as isolated and place it in thermal contact with a second body which
represents the outside world then the temperature of the first body will be equal to the 
second body, which acts as a heat reservoir, when thermal equilibrium is achieved. This
is the second step in Boltzmann's argument. In any event, extensivity (additivity) is
built into Boltzmann's principle through the logarithmic relation between entropy
and probability.\par
Finally, to rely completely on Boltzmann's principle to construct a complete
statistical mechanics is inadequate since Boltzmann's principle provides only part of the
probability distribution. On account of the logarithmic dependence 
(\ref{eq:Boltzmann}), this probability distribution turns out to be an error law for 
an extensive variable which equates
the most probable value of this quantity to the arithmetic mean of the measurements 
\cite{Keynes,Lavenda91}.

\par

\end{document}